\documentclass[reprint,aps,prl,showpacs]{revtex4-1}
\usepackage{graphicx,epsfig}
\usepackage{epstopdf}
\usepackage{bm}
\usepackage{amsmath}
  
\begin{document}

\title{Coexistence of quantum and classical flows in quantum turbulence in the $T=0$ limit}

\author{P. M. Walmsley and A. I. Golov}

\affiliation{School of Physics and Astronomy, The University of Manchester, Manchester M13 9PL, UK}

\date{\today}

\begin{abstract}
Tangles of quantized vortex line of initial density ${\cal L}(0) \sim 6\times 10^3$\,cm$^{-2}$ and variable amplitude of fluctuations of flow velocity $U(0)$ at the largest length scale were generated in superfluid $^4$He at $T=0.17$\,K, and their free decay ${\cal L}(t)$ was measured. If $U(0)$ is small, the excess random component of vortex line length firstly decays as ${\cal L} \propto t^{-1}$ until it becomes comparable with the structured component responsible for the classical velocity field, and the decay changes to ${\cal L} \propto t^{-3/2}$. The latter regime always ultimately prevails, provided the classical description of $U$ holds.
A quantitative model of coexisting cascades of quantum and classical energies describes all regimes of the decay. 
\end{abstract}

\pacs{67.25.dk, 47.27.wg}
\maketitle

Turbulent flow in classical fluids can be described \cite{Farge2001} as a superposition of coherent vortices, possessing large non-equilibrium energy and responsible for the cascade of this energy towards smaller length scales, and a random incoherent flow, which is in equilibrium and dissipates at the smallest scales by viscosity. The energy spectrum, i.\,e. contributions from velocity fluctuations at different length scales, adjusts self-consistently to maintain the continuity of the cascade's energy flux \cite{K41,Legacy}.

Vortices in superfluid $^4$He are different \cite{DonnellyBook} in that all of them have fliamentary cores surrounded by inviscid flow of identical velocity circulation $\kappa = h/m = 0.997\times 10^{-3}$\,cm$^{2}$\,s$^{-1}$ ($h$ and $m$ being the Plank's constant and atomic mass of $^4$He, respectively) \cite{DonnellyBook}. Hence, turbulence in this system ({\it quantum turbulence} or QT) is a tangle of vortex lines \cite{Feynman1955,VinenReview}. 
Yet, QT might be an analog of the classical scenario in that there are two coexisting structures \cite{BaggaleyPRL2012}: one (flow round bundles of vortex lines) possesses all properties of the classical coherent vortices, while another incoherent component (flow round individual lines) is responsible for the transfer of energy towards the dissipative processes at smaller scales and adjusts its own extent self-consistently. Importantly, the concept of the energy cascade is still potent \cite{Svistunov95}.

Of special interest is the limit of zero temperature, $T=0$, at which QT is non-dissipative down to length scales much smaller than the typical distance between vortices, $\ell = {\cal L}^{-1/2}$,  where ${\cal L}$ is the length of vortex line per unit volume. The nature and rate of the corresponding energy cascade and ultimate dissipative processes remain open questions of fundamental importance \cite{RingsNemirovsky,LaurieJLTP2015}. 
Two extreme cases of QT in the $T=0$ limit have been studied \cite{WalmsleyPRL2008,QT0,ZmeevPRL2015} and revealed different types of free decay ${\cal L}(t)$. 
One (`ultraquantum' or `Vinen QT') is a random tangle of vortex lines with negligible  velocity fluctuations at length scales $r \gg \ell$. Such a tangle is fully described by ${\cal L}$. 
Another limit (`quasiclassical' or `Kolmogorov QT') is that of partially polarized tangles with the dominant contribution to energy coming from flow round many vortex lines. 
Here a second parameter is required, the amplitude of velocity fluctuations $U$ at the integral length scale $L_i$ -- usually of order the container size $D$. 

Many questions remained. Which of these regimes is transient and which is the ultimate `equilibrium' type? Which parameters describe their interplay? What is the ratio of the contributions from the coherent and random components to the vortex length in the `equilibrium' state? Our experiment, in which vortex tangles were generated with a known value of $U$, answers these questions. 

The energy, per unit mass, of the turbulent state is the volume-averaged ${\cal E} = \frac{1}{2}<v^2>$ with velocity $v$ given by the Biot-Savart integral over all vortex lines. We consider a developed bulk QT, for which $L_i \gg \ell$. Then there are two major contributions to the energy, ${\cal E}={\cal E}_q + {\cal E}_c$. In the near field $r \ll \ell$, the `quantum energy' is dominated by the velocity of fluid circulating round individual lines, 
\begin{equation}
{\cal E}_q \approx \gamma \kappa^2 {\cal L},
\label{Eq}
\end{equation} 
where $\gamma \approx \ln(\ell/a_0)/4\pi$ and vortex core radius is $a_0 \approx 1.3$\,\AA\ \cite{DonnellyBook}. In our experiments, $\ell$ is within the range 0.14--2\,mm; hence, $\gamma \approx 1.2 \pm 0.1 \approx const$. 
  On the other hand, in the far field $r \gg \ell$ (`classical length scales'), the flow velocity arises from contributions of many aligned vortex lines. 
    If the forcing is at length scale $D$, and $Re_s \equiv \frac{UD}{\kappa} \gg 1$ \cite{VolovikQT}, then the coarse-grained velocity field should obey classical fluid dynamics with no dissipation. Hence, the Kolmogorov-like energy cascade \cite{K41} is expected, with the classical energy dominated by $U$,  
    \begin{equation}
{\cal E}_c \approx \frac{1}{2}U^2.
\label{Ec}
\end{equation} 
In the $T=0$ limit, the energy could be removed either by phonon emission due to short-wavelength Kelvin waves \cite{VinenKWaves,KSPhonons} or diffusion of small vortex rings  \cite{RingsBarenghi,RingsNemirovsky,WalmsleyPRF2016,LaurieJLTP2015,Yano}. Both processes are related to length scales $r \lesssim \ell$, and are fuelled by vortex reconnections \cite{NazarenkoReconnections,KSPRB2008}. 
The flux of energy towards these dissipative processes, is expected to obey \cite{Vinen1957,Vinen2000} 
\begin{equation}
\epsilon_d = \zeta \kappa^3 {\cal L}^2.
\label{zeta}
\end{equation}
Whether the dimensionless parameter $\zeta \sim 1$ depends on the tangle's polarization \cite{LNR2007} is still an open question.

With dominant flux of quantum energy, $|{\dot {\cal E}}_q| \gg |{\dot {\cal E}}_c|$, equating it to the dissipation rate, $-{\dot {\cal E}_q} = -\gamma \kappa^2 {\dot {\cal {L}}} = \epsilon_d$, results in the free decay of Vinen QT: 
 \begin{equation}
{\cal L} = \frac{\gamma}{\zeta \kappa}(t+t_V)^{-1},
\label{L-t-Vinen}
\end{equation}
with $t_V = \frac{\gamma}{\zeta\kappa{\cal L}_0}$ (where ${\cal L}_0 = {\cal L}(0)$). Such decay with a universal prefactor, corresponding to $\zeta \approx 0.10$, was observed in QT generated after a brief injection of ions in cells of different sizes \cite{WalmsleyPRL2008, ZmeevPRL2015} and also in numerical simulations of Vinen QT \cite{TsubotaPRB2000,KondaurovaVinenQT}.

In the opposite limit of Kolmogorov QT with dominant flux of classical energy, $|{\dot {\cal E}}_c| \gg |{\dot {\cal E}}_q|$, the rate of the energy release is controlled by the lifetime of largest eddies and the cascade time, both of order $L_i/U$. We hence assume that ${\dot {\cal E}}_c \sim -U^3/L_i$ and the energy flux at smallest lengths $\epsilon_c \sim U^3/L_i$ \cite{ClassicalFlux}. For constant $L_i \sim D$,
\begin{equation}
\epsilon_c =  \beta D^2 (t+t_K)^{-3}, 
\label{epsilon_c}
\end{equation}
with $t_K = a\frac{D}{U_0}$ (where $U_0 = U(0)$) and the prefators $a,\beta \sim 1$ depending on the container shape and boundary conditions \cite{ZmeevPRL2015}. Equating $\epsilon_c = \epsilon_d$ results in  
 \begin{equation}
{\cal L} = \left(\frac{\beta}{\zeta}\right)^{1/2} \kappa^{-3/2} D (t+t_K)^{-3/2},
\label{L-t-Kolmogorov}
\end{equation}
typical for decaying QT with the classical inertial length saturated by the container size \cite{SmithWallBounded}.
Such decay with $\zeta \sim 0.1$ 
was observed for QT in the $T=0$ limit, generated either by a towed grid or after a long intensive injection of ions \cite{ZmeevPRL2015}. 

In the present work, we developed a method of generating QT, in which $U$ can be controlled. 
A cubic volume with sides $D=4.5$\,cm, made of six earthed metal plates, contained $^4$He with $^3$He fraction $2\times 10^{-11}$ \cite{purity2} at pressure 0.1\,bar. Experiments were conducted at temperature $T=0.17$\,K, at which the normal fraction $\rho_n/\rho = 1.0\times 10^{-7}$ \cite{DatasheetDonnellyBarenghi1998} and mutual friction parameter $\alpha = 8\times 10^{-9}$ \cite{KS_MF} are negligible. The mean density of vortex lines ${\cal L}$ in the cell was evaluated by measuring the losses of charged vortex rings (CVRs) 
propagating from an injector in the centre of a side plate to the collector at the opposite side \cite{CVR_technique}. 
In order not to affect the decaying QT by the injected CVRs, each realization of vortex tangle, decaying for time $t$ after turning off the injection, was only probed once.

\begin{figure}
\includegraphics[width=8.4cm]{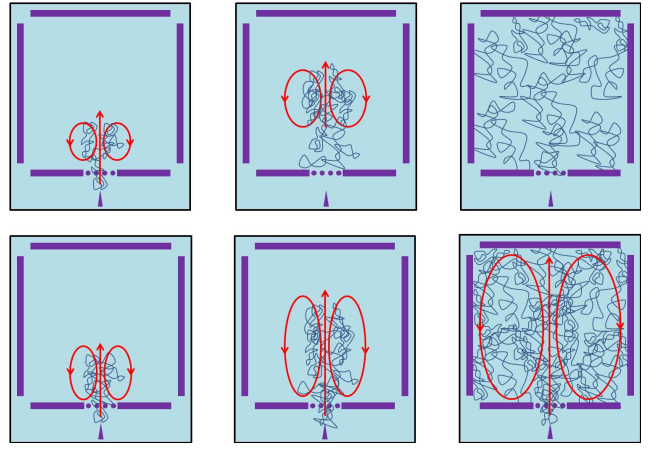}
\caption{Side cross-section of the experimental cell with the injector tip and grid at the bottom. The top row, left to right, illustrates the development of the vortex tangle (blue) and large-scale flow (red) after a brief injection. The bottom row shows the development during a continuous injection. }
\label{FigCell}
\end{figure}

The turbulence was generated by an injector of electrons in the middle of the bottom plate. This had a field-emission tip, to which a negative voltage of magnitude $V_*$ in the range 290--380\,V was applied  during time $\Delta t_*$ between 1\,s and 500\,s, resulting in the current of magnitude $I_*(V_*)$ in the range 0.8--470\,pA to a grid 2\,mm from the tip. Injected electrons, each in a bubble of radius 2\,nm \cite{bubble} (`negative ions'), immediately nucleate small vortex rings which quickly (during first $\sim 0.15$\,s) build up a dense vortex tangle between the tip and the grid \cite{WalmsleyPRB_rotatingQT}. The ions remain trapped on vortex lines until they reach the grid where most of them terminate, while the jet of fluid continues into the cell. Thus, by exerting force on these ions, the turbulence is simultaneously forced both on small lengths $\sim \ell$ due to the ballooning out of the charged vortex segments leading to the growth of the line density $\cal L$, and on large scales $\sim D$ due to the increase in the mean velocity $\sim U$ of the jet. 

We relate $U$ to the total hydrodynamic impulse $\cal P$ through ${\cal P} \sim \rho D^3 U$, where $\rho = 145$\,kg\,m$^{-3}$ is the density of helium. Before reaching the grid, each ion transfers to the fluid impulse $eV_*/v_*$ (with $v_* \sim 0.2$\,m\,s$^{-1}$ being the mean velocity of ions dragged by electric field through the slower vortex tangle as a consequence of frequent reconnections when at $T<0.7$\,K \cite{McClintock_charged_tangle, JLTP_charged_tangle}).
The rate of transfer of impulse to the jet into the cell (see Fig.\,\ref{FigCell}) is hence $\dot{{\cal P}_+} \approx V_*I_*/v_*$, while the rate of loss is $\dot{{\cal P}_-} \sim -{\cal P}/\tau_{\cal P} \sim - \frac{{\cal P}^2}{\rho D^4}$, where $\tau_{\cal P} \sim D/U$ is the time required for the jet to reach the opposite wall, during which the impulse is conserved. 
The dependence $U(t)$ during injection, which commenced at $t=-\Delta t_*$ and ended at $t=0$, can be found from the solution of the equation ${\dot {\cal P}} = {\dot {\cal P}}_+ + {\dot {\cal P}}_-$: 
\begin{equation}
U \sim \frac{D}{\tau_*} \tanh \left(\frac{t+\Delta t_*}{\tau_*}\right),
\label{Tau_U}
\end{equation}
where $\tau_* = D^2\left( \frac{\rho v_*}{V_*I_*} \right)^{1/2}$ is the time scale for the given injection intensity $I_*(V_*)$, that separates regimes of growing $U(t)$ and saturated $U$.  
In what follows, we will need a general expression for the value of $t_K$ in (\ref{epsilon_c}), where $U_0 = U(V_*, I_*)$ at $t=0$ (here $a,b \sim 1$),
\begin{equation}
t_K = a\tau_* \left[ \tanh \left(\frac{\Delta t_*}{b\tau_*}\right) \right]^{-1}.
\label{t2}
\end{equation}

 \begin{figure}
\includegraphics[width=8cm]{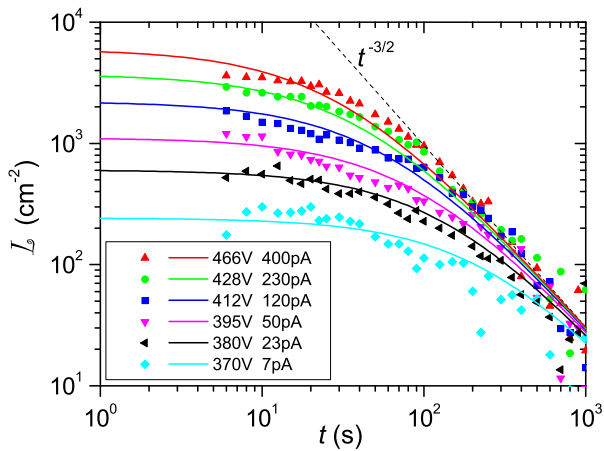}
\caption{${\cal L}(t)$ for decaying tangles, forced for the same duration $\Delta t_* = 300$\,s but by different $V_*$ and $I_*$ (listed in legend). Lines correspond to Eq.\,\ref{L-t-Kolmogorov} (see text).}
\label{FigLtLongInj}
\end{figure}

This relation was firstly tested for the limit of long injection, $\Delta t_* \gg \tau_*$, in which 
 turbulence with a steady classical energy flux is established. 
In Fig.\,\ref{FigLtLongInj} we plot experimental ${\cal L}(t)$ for several decaying vortex tangles generated by injections of the same duration $\Delta t_* = 300$\,s but of different intensities $(V_*, I_*)$ for which $\tau_*$ takes values from  210\,s to 25\,s. The solid lines are Eq.\,\ref{L-t-Kolmogorov} with the initial values corresponding to $t_K = a\tau_* =  a D^2 \left( \frac{\rho v_*}{V_*I_*} \right)^{1/2}$ (from Eq.\,\ref{t2} in the $\Delta t_* \gg \tau_*$ limit) with $a=1.2$, and the common late-time asymptotic with $(\beta/\zeta)^{1/2} = 7$.  
We thus confirm that a long intensive injection can generate Kolmogorov turbulence whose decay follows Eq.\,\ref{L-t-Kolmogorov}, and that our model for the amplitude of injected large-scale velocity $U(V_*,I_*,\Delta t_*)$ and associated time scale $t_K$, Eq.\,\ref{t2}, is in agreement with experiment.

 \begin{figure}
\includegraphics[width=8cm]{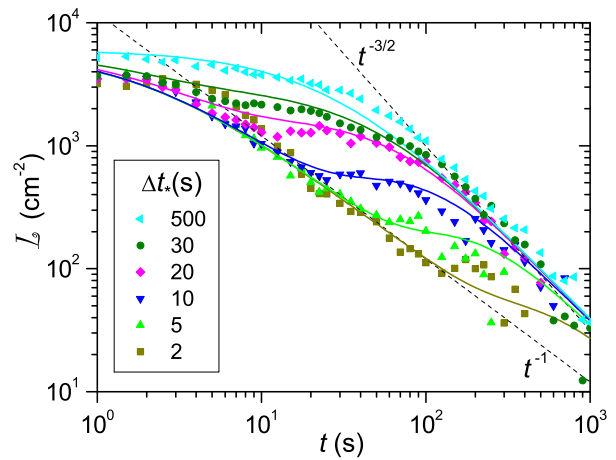}
\caption{${\cal L}(t)$ for decaying tangles forced by the same $I_* = 466$\,pA at $V_* = 380$\,V (making $\tau_* = 18$\,s), but for different durations $\Delta t_*$. Dashed lines  correspond to Eq.\,\ref{L-t-Vinen} and Eq.\,\ref{L-t-Kolmogorov} with $t_V = t_K=0$. Solid lines are solutions of Eq.\,\ref{Master-2} with ${\cal L}_{0} = 6\times 10^{3}$\,cm$^{-2}$ but different $t_K(V_*,I_*,\Delta t_*)$ (see text). 
}
\label{FigLtVarDeltaT}
\end{figure}

In Fig.\,\ref{FigLtVarDeltaT}, which is the main result, we show the measured ${\cal L}(t)$ for several decaying vortex tangles, created with different initial values of $U_0$ by varying $\Delta t_*$ while keeping $V_*$ and $I_*$ the same. Except for the top dataset with the longest $\Delta t_* = 500$\,s, the decay begins with a universal dependence ${\cal L} \propto (t+t_V)^{-1}$, expected for Vinen QT (\ref{L-t-Vinen}). This dependence is continued for some time until it gradually switches to ${\cal L} \propto (t+t_K)^{-3/2}$, characteristic of wall-bounded Kolmogorov QT (\ref{L-t-Kolmogorov}). The longer the injection time $\Delta t_*$ (i.\,e. the greater the value of $U_0$), the earlier the switch occurs.  
To model the dependence ${\cal L} (t)$ during free decay, we write the energy balance at length scales $\sim \ell$, 
\begin{equation}
\epsilon_c(U_0, t) - \dot{{\cal E}_q}({\cal L}) = \epsilon_d({\cal L}).
\label{Master}
\end{equation}
Following \cite{VinenL'vov2016} we assume that the flux of classical energy (\ref{epsilon_c}) effectively reaches this length scale after the delay time $\sim t_K(V_*, I_*, \Delta t_*)$ from the beginning of injection at $t=-\Delta t_*$. 
We hence introduce a simple delay function \cite{VinenL'vov2016} $F(t+ \Delta t_*,t_K)=(1-e^{-(t+\Delta t_*)/t_K})^2$, so $\epsilon_c = F(t+\Delta t_*, t_K) \beta D^2 (t+t_K)^{-3}$. 
Eq.\,\ref{Master} becomes,
\begin{equation}
\dot{\cal L} =  \frac{\beta D^2}{\gamma\kappa^2}\frac{(1-e^{-\frac{t+\Delta t_*}{t_K}})^2}{(t+t_K)^3} - \frac{\zeta \kappa}{\gamma} {\cal L}^2 ,
\label{Master-2}
\end{equation} 
which can be solved numerically for ${\cal L}(t)$ subject to the initial parameters ${\cal L}_0$ and $U_0$ (via $t_K = aD/U_0$).
 In Fig.\,\ref{FigLtVarDeltaT} we show solutions of Eq.\,\ref{Master-2} with $\zeta = 0.10$ and $\beta =  4.9$ (i.\,e. with the same late-time asymptotic (\ref{L-t-Kolmogorov}) with $\left(\frac{\beta}{\zeta}\right)^{1/2}=7$ as in Fig.\,\ref{FigLtLongInj}), with values of $t_K(V_*, I_*, \Delta t_*)$ calculated by Eq.\,\ref{t2} with $a=1.2$, $b=0.7$ and $v_*=0.2$\,m\,s$^{-1}$, and with one-for-all ${\cal L}_0= 6\times10^3$\,cm$^{-2}$. The good agreement with all experimental ${\cal L}(t)$ suggests that the model (\ref{Master-2}) adequately represents the dynamics of QT of arbitrary degree of polarization. 
We will now discuss some implications of the model.
 
 At early times, whether the decay will begin from either Vinen or Kolmogorov type depends on the interplay of the total ${\cal L}_0$ and the vortex length necessary to sustain the classical cascade ${\cal L}_{\parallel 0} = \left(\frac{\beta}{\zeta a^3\kappa^3 D}\right)^{1/2}U_0^{3/2}$ (from Eq.\,\ref{L-t-Kolmogorov}). If ${\cal L}_0 \sim {\cal L}_{\parallel 0}$, only the Kolmogorov decay (6) will be observed from the very begining (like the top dataset in Fig.\,\ref{FigLtVarDeltaT}). On the other hand, with ${\cal L}_0 \gg {\cal L}_{\parallel 0}$, i.\,e. 
$Re_s(0) \ll a(\frac{\zeta}{\beta})^{1/3} (D^2{\cal L}_0)^{2/3} \approx 0.3 (D^2{\cal L}_0)^{2/3}$, the Vinen regime (\ref{L-t-Vinen}) would firstly dominate.  
For $\ell \ll D$, any initially excessive quantum energy ${\cal E}_q$ always decays faster than the classical ${\cal E}_c$, because the decay time associated with the Vinen regime (\ref{L-t-Vinen}), $\tau_V \simeq \frac{\gamma}{\zeta\kappa{\cal L}}$, is shorter than that for the Kolmogorov regime  (\ref{L-t-Kolmogorov}), $\tau_K \simeq \left(\frac{\beta D^2}{\zeta \kappa^3 {\cal L}^2}\right)^{1/3}$. 
For the ultimate Kolmogorov decay ${\cal L} \propto t^{-3/2}$ to be restored while 
$\ell \ll D$, the condition is ${\cal L}(t_K) \gg D^{-2}$, i.\,e. 
$Re_s(0) \gg 2a\left(\frac{\zeta}{\beta}\right)^{1/3} \sim 1$ \cite{commentVolovik}. 
And in the opposite limit, $Re_s(0) \ll 1$, only the Vinen decay  could be observed. 
Note that this criterion differs from theory by Barenghi {\it et al.} \cite{Barenghi2016}.  
They claim that if a {\it spatially-uniform} injection of small vortex rings is stopped before the inverse cascade (which promotes large-scale velocity fluctuations upon the tangling of vortex rings) extends up to the largest length scale $\sim D$, only the ${\cal L} \propto t^{-1}$ decay can be observed. 
However, in all our experiments in which either a beam of vortex rings \cite{WalmsleyPRL2008} or a vortex tangle (this work) is injected, the large-scale velocity component $U_0$ is present from the very moment of tangling -- without the need of an inverse cascade. This is because of the {\it collimated} profile of resulting jets. 

Finally, approaching the crossing point of the asymptotics (\ref{L-t-Vinen}) and (\ref{L-t-Kolmogorov}) 
 at density ${\cal L}_c = \frac{\gamma^3}{\zeta^2\beta}D^{-2} \approx 36D^{-2} \approx 2$\,cm$^{-1}$, the formal solution of Eq.\,\ref{Master-2} deviates from (\ref{L-t-Kolmogorov}) and eventually switches to (\ref{L-t-Vinen}). However, the model of homogeneous QT might no longer be adequate at corresponding $\ell \sim D/6$. 
 Instead, it is expected that remnant vortices will replace the decaying tangle at similar densities  ${\cal L}_{r} \sim 2\ln(D/a_0)D^{-2} \sim 40 D^{-2}$ \cite{AwschalomPRL1984}.
  
 Let us turn to the question whether coherent bundles of vortex lines might be identifiable during the Kolmogorov decay ${\cal L} \propto t^{-3/2}$. This could be characterized by the ratio $\chi \equiv {\cal L}_{\parallel}/{\cal L}$, where ${\cal L}_{\parallel}$ is the length of aligned vortices which generate the quasiclassical velocity field, while the rest, ${\cal L}_{\times} = {\cal L} - {\cal L}_{\parallel}$, is made of random vortex segments. A similar decomposition was introduced previously \cite{Lipniacki, WalmsleyPRL2007, BarenghiRoche, BradleyPRL2008} and found meaningful \cite{BaggaleyPRL2012}. To estimate  ${\cal L}_{\parallel}$, we sum, in quadrature, contributions to classical vorticity from different length scales \cite{WalmsleyPRL2007}: 
\begin{equation}
\kappa^2 {\cal L}_{\parallel}^2 = \int_{\pi/D}^{x \pi/\ell}k^2E_k \approx \frac{3C}{4}x^{4/3}\pi^{4/3}\epsilon_c^{2/3}{\cal L}^{2/3},
\label{enstrophy} 
\end{equation}
where $E_k=C\epsilon_c^{2/3}k^{-5/3}$ is the Kolmogorov K41 spectrum with $C\approx 1.5$, and $x \sim 1$ defines the effective cut-off wavenumber for the classical spectrum.  
  If the classical energy flux  $\epsilon_c$ dominates, $\epsilon_c \approx \epsilon_d$, then, with Eq.\,\ref{zeta}, 
\begin{equation}
\chi \equiv \frac{{\cal L}_{\parallel}}{{\cal L}} \approx \frac{\sqrt{3C}}{2}(x^2\pi^2\zeta)^{1/3} \sim 1.
\label{LcL} 
\end{equation}
Thus, the late-time decaying tangles maintain a substantial and constant degree of alignment \cite{CommentChi}.

In fact, the phenomenological expression (\ref{zeta}) for the rate of dissipation $\epsilon_d$ might have alternatives. One could argue that the component ${\cal L}_{\times}$ is passively advected by classical flow and is hence involved in the transfer of energy at the same rate as in Vinen QT \cite{Lipniacki, BarenghiRoche}, while ${\cal L}_{\parallel}$ might not contribute to the removal of energy as efficiently because it is related to the classical velocity field which evolves at its own pace. 
Hence, as a special case,
 \begin{equation}
\epsilon_d \approx \zeta \kappa^3 {\cal L}_{\times}^2 = \zeta \kappa^3 ({\cal L}- {\cal L}_{\parallel})^2,
\label{zeta1}
\end{equation}
with $\zeta = 0.10$. Eq.\,\ref{zeta1} would still be compatible with all previous experimental observations, including those for grid turbulence \cite{ZmeevPRL2015}, provided ${\cal L}_{\times} \sim {\cal L}$. Assuming that, like in the previous case of Eq.\,\ref{LcL}, $\chi$ is constant during the late-time decay, 
 and using (\ref{enstrophy}) and (\ref{zeta1}), we arrive at 
\begin{equation}
\chi \approx \frac{\sqrt{3C}}{2}(x^2\pi^2\zeta)^{1/3}(1-\chi)^{2/3} \approx x^{2/3}(1-\chi)^{2/3}.
\label{LcL1} 
\end{equation}
Its solution for $x=1$ is $\chi=0.57$ -- indicating that 
 ${\cal L}_{\parallel}$ and ${\cal L}_{\times}$ are indeed comparable. 
 We solved Eq.\,\ref{Master} numerically with $\epsilon_d$ given by (\ref{enstrophy})\&(\ref{zeta1}), instead of (\ref{zeta}). It turned out, all experimental data ${\cal L}(t)$, shown in Fig.\,\ref{FigLtLongInj} and Fig.\,\ref{FigLtVarDeltaT}, can be modelled nearly as satisfactorily, e.\,g. if one chooses $\zeta = 0.10$, $\beta = 0.8$, $a=1.0$, $b=0.7$ and $x=0.5$. Thus, the important question, which of Eq.\,\ref{zeta} and Eq.\,\ref{zeta1} is more appropriate to describe the dynamics of QT of various degrees of polarization, requires further investigation.
 
To conclude, we developed a technique of generating, in the $T=0$ limit, QT with the known amplitude $U_0$ of flow velocity at the integral length $D$. For the range of injection conditions as in Fig.\,\ref{FigLtVarDeltaT}, the superfluid Reynolds number $Re_s$ spans the range between 25 and 650. Our model, which combines the fluxes of quantum and classical energy \cite{WalmsleyPRL2008},
describes all features of the observed decays ${\cal L}(t)$. 
If the initial line density ${\cal L}_0$ greatly exceeds the aligned fraction ${\cal L}_{\parallel 0}(U_0)$, associated with the quasiclassical flow, ${\cal L}$ rapidly decreases following a universal decay law of Vinen QT, ${\cal L} \propto (t+t_V)^{-1}$. Yet, the initial quasiclassical flow decays slower, and when ${\cal L}_{\parallel}/{\cal L}$ reaches $\sim 1$, the late-time decay of Kolmogorov QT ${\cal L} \propto D(t+t_K)^{-3/2}$, universal for the given container, is maintained. 
Only for very small initial $U_0 \lesssim \kappa/D$ (i.\,e. when $Re_s \lesssim 1$ is too small to warrant classical behavior of even largest eddies), can this ultimate regime never be reached.

\begin{acknowledgments}
We acknowledge fruitful discussions with Joe Vinen and Henry Hall, help by Alexandr Levchenko and Steve May in constructing equipment, and supply by Peter McClintock of isotopically-pure $^4$He. Support was provided by EPSRC under EP/E001009, GR/R94855, EP/I003738/1, and EP/H04762X.
\end{acknowledgments} 

\end{document}